\documentstyle[12pt]{article}
\if@twoside
 \else
 \fi

 \parskip \medskipamount

\begin{document}
\hfill{LA-UR-93-1062}

\vspace{7pt}
\begin{center}
{\large\sc{\bf Bosonization in 2+1 dimensions
without Chern - Simons attached.}}

\baselineskip=12pt
\vspace{35pt}

A. Kovner$^ *$ and P. S. Kurzepa $^{**}$\\
\vspace{10pt}
Theory Division, T-8,
Los Alamos National Laboratory,
MS B-285\\
Los Alamos, NM 87545\\
\vspace{55pt}
\end{center}
\begin{abstract}
We perform the complete bosonization of 2+1
dimensional QED with one fermionic flavor in the
Hamiltonian formalism. The fermion operators are
explicitly constructed in terms of the vector
potential and the electric field. We carefully specify
the regularization procedure
involved in the definition of these operators, and calculate
the fermionic bilinears and the energy - momentum tensor.
The algebra of bilinears exhibits the Schwinger terms which
also appear in perturbation theory. The bosonic Hamiltonian density
is a local polynomial function of $A_i$ and $E_i$, and we
check explicitly the Lorentz invariance of the resulting bosonic theory.
Our construction is conceptually very similar to Mandelstam's
construction in 1+1 dimensions, and is dissimilar from the recent
bosonization attempts
in 2+1 dimensions which hinge crucially on the existence
of a Chern - Simons term.
\end{abstract}

\vspace{50pt}
\
\newline
*KOVNER@PION.LANL.GOV
\newline
**KURZEPA@MUON.LANL.GOV
\vfill
\pagebreak
It is well known that, in 1+1 dimensions, fermion field
operators can be constructed in terms of local bosonic fields.
This bosonization procedure is of great theoretical
interest, since it provides a method of mapping different local
quantum field theories with a priori different Hilbert spaces onto each other.
This mapping has proved to be very helpful in the analysis
of different 1+1 dimensional models. After the pioneering
works of Coleman \cite{coleman} and Mandelstam \cite{mandelstam},
bosonization (and its nonabelian
generalization \cite{witten})
has been used in a variety of contexts: relativistic quantum
field theory, condensed matter systems, string theory.

Clearly, the extension of bosonization to higher
dimensions is highly desirable.
There were several early attempts \cite{earlybos}, but
those did not lead to local bosonic theories. Until recently
it has been tacitly assumed that such an extension is
impossible. The feeling was that 1+1 dimensions is a very special
case since there is no spin in 1+1 dimensions, and therefore
no ''real'' difference between bosons and fermions.

The renewed interest in this problem was triggered by a
possible connection between the phenomenon of
Fermi - Bose transmutation in 2+1 dimensions and
novel condensed matter
systems, most notably high $T_c$ superconductors.
Polyakov argued \cite{polyakov} that the addition of
a Chern - Simons term \cite{deser} to the Lagrangian of
QED$_3$ changes the statistics of charged excitations. His
argument was elaborated further in many works
\cite{kurzepa}, and the explicit realization of this
Chern - Simons induced mechanism of Fermi - Bose
transmutation in lattice theories
has been given in \cite{fradkin}, \cite{luscher}.

There is, however, an important conceptual difference
between the Mandelstam construction in 1+1 dimensions
and the Chern - Simons bosonization in 2+1dimensions. In the former
case bosonization is achieved essentially through a duality
transformation: one constructs fermionic operators
$\psi$, which carry a global U(1) fermion number charge,
in terms of a local bosonic field $\phi$,  which is
itself neutral. The U(1) charge, when expressed in terms
of $\phi$, is a topological, rather than a Noether,
charge. In the 2+1 - dimensional construction described
above the fermionic field $\psi$ is constructed in terms
of a bose field $\phi$, which carries the same global quantum
numbers. The element of duality is notably missing here.
Moreover, in this picture, a free fermion is bosonized into
a particle interacting with a vector potential with
the Chern - Simons action. This bose field is not gauge
invariant, and after gauge fixing becomes nonlocal. So one
does not really construct
fermionic operators in terms of {\it local} bosonic fields.

The 2+1 dimensional construction depends crucially
on the existence of the  Chern - Simons term. The Fermi - Bose
transmutation occurs only for a fixed value of its coefficient.
For other values of this coefficient, the procedure gives
nonlocal anyonic fields with fractional statistics.
Another unsatisfactory feature of the construction is that it
has been consistently implemented only for nonrelativistic
fermions \cite{fradkin}, \cite{luscher}.
When applied to continuum relativistically invariant
theories \cite{semenoff} it suffers from regularization ambiguities,
and it is not clear how to interpret the formal results.
In fact, the covariant Dirac field has not yet been constructed, even formally,
in this framework.

In this letter we take a different approach to bosonization. It is
the
direct extension of Mandelstam's procedure to higher dimensions.
Our aim is to construct a Dirac doublet of fermionic operators
in 2+1 dimensions in terms of {\it local} bosonic fields. As a working
hypothesis we assume that, in terms of the bose fields,
the fermion number charge is topological,
and thus the associated current is trivially conserved. We are
not aware of any theorem that states this, but such is the case
in all of the known examples where either exact bosonization can be performed
(in 1+1 dimensions) \cite{coleman}, \cite{mandelstam}, or fermionic
excitations
are known to exist in the spectrum of a local bosonic theory in
2+1 \cite{ijmp}, or 3+1 \cite{zahed} dimensions.

Given this assumption, the theory that suggests itself as a convenient
candidate for bosonization is QED with one two - component
Dirac fermion. The reason is that Maxwell's equations,
\begin{equation}
J^\nu=\frac{1}{e}\partial_\mu F^{\mu\nu}
\end{equation}
tell us that, if one takes as basic variables the electric field $E_i$
and the vector potential $A_i$, fermion number is trivially
conserved, due to  the antisymmetry of $F_{\mu\nu}$. The fermion
number charge is topological,
since it is
equal to the surface integral of $E_i$  {\it at spatial infinity}.
Since there are
no additional flavor symmetries here, it is
the simplest  model of its kind.

Our strategy is as follows.
Consider the Hamiltonian of the theory in the temporal (Weyl) gauge $A_0 = 0$,
\begin{equation}
H=\frac{1}{2}E^2+\frac{1}{2}B^2+\bar\psi\gamma^i(i\partial_i+eA_i)\psi+m\bar\psi\psi
\label{ham}
\end{equation}
together with Gauss's constraint,
\begin{equation}
\partial_iE_i=e\psi^\dagger\psi
\label{const}
\end{equation}
We will solve the constraint by constructing
the doublet of anticommuting fermionic operators $\psi_\alpha$
in terms of the local fields $E_i$ and their canonical conjugate
momenta $A_i$, $[E^i(x),A^j(y)]=i\delta^{ij}\delta^2(x-y)$.
Substituting those back into the Hamiltonian,
eq.(\ref{ham}), we will obtain the completely bosonized form of QED$_3$,
defined on the physical Hilbert space.
We will then calculate the fermionic bilinears, and check, that
they satisfy the tree level algebra, modified by Schwinger terms.
The appearance of these Schwinger terms is also
seen in perturbation theory at the one loop level.

To construct the operator $\psi$ we must first fix the gauge freedom
associated with the time independent gauge transformations, generated
by the Gauss's constraint. We do this by considering $\psi$ in the
Coulomb gauge. When written in terms of these operators, the
covariant derivative in eq.(\ref{ham})
contains only the transverse part of the vector potential $A_i$.
Furthermore, we will work with the gauge - invariant operators,
\begin{equation}
\psi^{CG}_\alpha(x)=\psi_\alpha(x)e^{ie\int d^2y e_i(y - x)A_i(y)}
\label{cg}
\end{equation}
where,
\begin{equation}
e_i(x)=-\frac{1}{2\pi}\frac{x_i}{x^2}
\end{equation}
is the electric field of a point
charge. (We will drop the superscript $CG$ from now on).
In addition to solving the constraint, eq.(\ref{const}), the
fermionic operators $\psi_\alpha$ must satisfy the following
conditions\footnote{Bosonization of the massive Schwinger model
in 1+1 dimensions can be formulated in precisely the same terms.}:

i. Carry unit electric charge,
$[\psi_\alpha(x),\partial_iE_i(y)]=e\psi_\alpha(x)\delta^2(x-y)$;

ii. Transform correctly under rotations. For convenience we
choose the basis of Dirac matrices in which the rotation
generator is diagonal,
$\gamma^0=\sigma^3,\  \gamma^1=i\sigma^2,\ \ \gamma^2=-i\sigma^1$. Then,
\begin{equation}
\psi_1\rightarrow e^{i\phi/2}\psi_1; \ \ \psi_2\rightarrow e^{-i\phi/2}\psi_2
\label{rot}
\end{equation}
 where $\phi$ is the rotation angle;

iii. Fermionic bilinears must be local operators, i.e.,
$[\psi^\dagger_\alpha\psi_\beta(x),O(y) ]=0$ for $ x\ne y$ and
for any local gauge invariant operator $O(x)$.

To satisfy the first condition, we take the following ansatz,
\begin{equation}
\psi_\alpha(x)=k\Lambda V_\alpha(x)e^{ie\int d^2y e_i(y -x)A_i(y)}U_\alpha(x)
\label{an1}
\end{equation}
where $\Lambda$
is the ultraviolet cutoff, $k$ is a constant, and the operators $V_\alpha$
and $U_\alpha$ commute with the charge density creation operator.
The following form of these operators (which is not final!) ensures
the anticommutativity of  the $\psi$'s,
and also has correct rotational properties,
\begin{eqnarray}
V_1(x)=-ie^{\frac{i}{2e}\int d^2y(\theta(x-y)-\pi)\partial_iE_i(y)}&;&
V_2(x)=-i V^\dagger_1(x) \\ \nonumber
U_1(x)=e^{-\frac{i}{2e}\int d^2y\theta(y-x)\partial_iE_i(y)}&;&
U_2(x)=U^\dagger_1(x)
\end{eqnarray}
Here $\theta(x)$ is the polar angle of the
point $x$ (for the precise definition of $\theta(x)$ see \cite{pap}).
With these definitions we find,
\begin{eqnarray}
\psi_1(x)\psi_1(y)&=&\psi_1(y)\psi_1(x)
e^{-i(\theta(y-x)-\theta(x-y))}=-\psi_1(y)\psi_1(x) \\ \nonumber
\psi_1(x)\psi^\dagger_1(y)&=&\psi^\dagger_1(y)\psi_1(x)
e^{i(\theta(y-x)-\theta(x-y))}=-\psi^\dagger_1(y)\psi_1(x) \\ \nonumber
\psi_2(x)\psi_2(y)&=&\psi_2(y)\psi_2(x)e^{i(\theta(y-x)-\theta(x-y))}=
-\psi_2(y)\psi_2(x) \\ \nonumber
\psi_2(x)\psi^\dagger_2(y)&=&\psi^\dagger_2(y)\psi_2(x)
e^{-i(\theta(y-x)-\theta(x-y))}=-\psi^\dagger_2(y)\psi_2(x) \\ \nonumber
\psi_1(x)\psi^\dagger_2(y)&=&\psi^\dagger_2(y)\psi_1(x)e^{-i\pi}=
-\psi^\dagger_2(y)\psi_1(x) \\ \nonumber
\psi_1(x)\psi_2(y)&=&\psi_2(y)\psi_1(x)e^{i\pi}=-\psi_2(y)\psi_1(x)
\end{eqnarray}

The factor $k\Lambda$ in eq.(\ref{an1}) takes care of the
correct dimensionality of the fermionic fields, and ensures
the correct normalization of the anticommutators,
\begin{equation}
\{\psi^\dagger_1(x),\psi_1(y)\}=\{\psi^\dagger_2(x),\psi_2(y)\}=\delta^2(x-y)
\end{equation}

However, the above manipulations are formal, since the commutators of V
and U with the $A$ - dependent factor in eq.(\ref{an1}) are singular.
The expression for the fermionic operators needs therefore
regularization. We shall regularize it by point splitting.
We also modify the expressions for $V(x)$ and $U(x)$, so
that they become
creation and annihilation operators of a magnetic vortex of
half - integer strength \cite{npb},
\begin{equation}
V_1(x)=-i\exp\left\{\frac{i}{2e}\int
d^2y\left[(\theta(x-y)-\pi)\partial_iE_i(y)
+2\pi G^{(2)}(y-x)\epsilon_{ij}\partial_iE_j(y)\right]\right\};
\label{vu}
\end{equation}
$$U_1(x)=
\exp\left\{-\frac{i}{2e}\int d^2y\left[\theta(y-x)\partial_iE_i(y)+
2\pi G^{(2)}(y-x)\epsilon_{ij}\partial_iE_j(y)\right]\right\}; $$
$$V_2(x)=-iV^\dagger_1(x); U_2(x)=U^\dagger_1(x)$$
Here $G^{(2)}(x-y)=-\frac{1}{4\pi}\ln( x^2\mu^2)$ is the two - dimensional
Green's function of the free massless field with IR cutoff $\mu$.
The physical interpretation of the operators $V_\alpha$ and $U_\alpha$ is
clear from the following commutation relations,
\begin{equation}
[V_1(x),B(y)]=-\frac{\pi}{e}V_1(x)\delta^2(x-y) ;\ \
[U_1(x),B(y)]=\frac{\pi}{e}U_1(x)\delta^2(x-y)
\end{equation}
The regularization we employ does not change either the anticommutation
relations or the rotational properties of the operators, but is
crucial to ensure the locality of the bosonized forms of the fermionic
bilinears.

The regularized expression for the fermionic operators involves an
average over the direction of the regulator $\eta$, which implements the
point splitting,
\begin{equation}
\psi_1(x)=\lim_{\Lambda\rightarrow \infty}\frac{k\Lambda}{2\pi }
\int d\hat\eta e^{-i\frac{\theta(\eta)}{2}}\psi_1^{\eta}(x); \ \
\psi_2(x)=\lim_{\Lambda\rightarrow \infty}\frac{k\Lambda}{2\pi }
\int d\hat\eta e^{i\frac{\theta(\eta)}{2}}\psi_2^{\eta}(x)
\label{psi}
\end{equation}
where
\begin{equation}
\psi^\eta_1(x)=
V_1(x+\eta)e^{ie\int d^2y e_i(y - x)A_i(y)}U_1(x-\eta);
\label{psieta}
\end{equation}
$$\psi^\eta_2(x)= V_2(x+\eta)e^{ie\int d^2y e_i(y - x)A_i(y)}U_2(x-\eta)$$

The phase factors introduced into the averages ensure the correct
rotational properties of the fermionic operators.
The length of the regulator $\eta$ is taken to be proportional to the
inverse of the UV cutoff, $|\eta|\propto 1/\Lambda$, the integral is
over the angle of the vector $\eta$, and $\hat\eta$ is the unit vector
in the $\eta$ direction. It can be explicitly checked that,
\begin{equation}
\{\psi_\alpha(x),\psi_\beta(y)\}=0, \ \ |x-y|>>1/\Lambda
\end{equation}
Therefore, in the limit $\Lambda\rightarrow \infty$, we regain the
standard anticommutation relations.
Equations (\ref{vu}), (\ref{psi}) and (\ref{psieta}) are our
final expressions for
the fermionic operators in terms of the bose fields $E_i$ and $A_i$.

 We next calculate the fermionic bilinears. Those should also
be defined using the point splitting procedure. The following
definition of the bilinears is a natural one,
\begin{equation}
J_\Gamma(x)=\bar\psi(x)\Gamma\psi(x)\equiv \frac{1}{8\pi}
\int d\hat\epsilon e^{i\chi_\Gamma(\hat\epsilon)}\left\{\left[
\psi^\dagger(x+\epsilon),\gamma^0\Gamma\psi(x-\epsilon)\right]
,e^{ie\int_{x-\epsilon}^{x+\epsilon}dx_i
A_i^{tr}}\right\}_{|\epsilon|\propto|\eta|}
\label{bilinear}
\end{equation}
It is implicitly understood in eq.(\ref{bilinear}) that the
limit $\Lambda\rightarrow \infty$ is taken after
(independent) averaging over the directions of $\epsilon$ and $\eta$.
The insertion of the Wilson factor is appropriate for the definition
of bilinears in a gauge theory. Since we are constructing the
Coulomb gauge fermions it is only the transverse part of the vector
potential, $A^{tr}$, that appears in the Wilson factor. The phase
$\exp\{i\chi_\Gamma\}$ is inserted to project onto
the relevant irreducible representation of the 2D
rotation group while averaging over $\hat \epsilon$.
Thus for $\Gamma=\gamma_0$ and $\Gamma=1$ we have
$\chi(\epsilon)=0$, while for $\Gamma=\gamma_+\equiv\gamma_1+i\gamma_2$,
$\chi(\epsilon)=\theta(\hat\epsilon)$ and for $\Gamma=\gamma_
-\equiv\gamma_1-i\gamma_2$, $\chi(\epsilon)=-\theta(\hat\epsilon)$.
The ratio $\eta/\epsilon$ is arbitrary, but the final results should not
depend on it. We also normal - order the bilinears with respect to the
perturbative vacuum.

We have calculated the bilinears using the definitions
of the fermionic operators given above. The calculation is performed,
in analogy to the 1+1 dimensional case, by expanding all quantities
in powers of the inverse
cutoff. This procedure has a certain caveat that one has to keep in mind.
The operators which are multiplied by inverse powers of $\Lambda$, and
are formally small, may, in fact, give finite contribution in the continuum
limit, if these operators have high enough dimensions. This problem
arises also in 1+1 dimensions, where the effect of the higher order terms is
to renormalize the coefficients of the lower dimensional operators in a way
consistent with the symmetries of the problem \cite{mandelstam} \footnote{As
in $1+1$ dimensions, this renormalization ambiguity affects only the spatial
components of the current.}.
Furthermore, in order to perform the actual calculations, we must control the
ultraviolet behavior of the bosonized theory.
This behavior is different from the
naive one, obtained in perturbation theory. This is related to the fact
that, apart from the
ultraviolet cutoff $\Lambda$, the bosonized
theory has an additional ultraviolet scale $\mu=(e^2\Lambda^2)^{1/3}$.
The appearance of this scale can be seen as follows.
The fermionic operator defined in eq.(\ref{psi}) is essentially
a product of a vortex and
an antivortex operator, separated by $\Lambda^{-1}$.
The vortex operator scales, at short distances,
as an exponential,
\begin{equation}
<V(x)V^*(y)>\propto|x-y|^\alpha
exp{\left [\frac{c}{e^2}|x-y|^{2 - \alpha }\right ]}
\end{equation}
where $c$ is a constant, and $\alpha /2$ is the scaling dimension of the
electric field, $<E(x)E(y)>\propto \frac{1}{|x-y|^\alpha }$.
Assuming the perturbative scaling of the electric field
at short distances, $\alpha = 3$,  we find that
the fermion operator that we have constructed, at distances
larger than $1/\Lambda$, but still small relative to any physical distance
scale, behaves, {\it roughly}, as,
\begin{equation}
<\psi^\dagger_\eta(x)\psi_\xi(y)>\propto|x-y|^\gamma
exp{\left [\frac{c\hat\eta\hat\xi}{e^2\Lambda^2
|x-y|^3}\right ]}
\label{psicor}
\end{equation}
The scale $\mu$ is therefore clearly a {\it crossover scale}.
At distances larger than,
\begin{equation}
|x-y|^3\propto \frac{1}{e^2\Lambda^2}\equiv\frac{1}{\mu^3}
\label{dist}
\end{equation}
the exponential in eq.(\ref{psicor}) can be expanded in power series.
In that case
the fermion propagator scales as a power, which is consistent
with its perturbative behavior.

At distances smaller than $1/\mu$
the non - point - likeness of $\psi$ becomes important.
As a result the  scaling behavior at these short distances
in the bosonized theory must be different
from the perturbative one.
In fact, if the fermionic operator is still to scale with a power law
for $\mu<<1/x<<\Lambda$, the leading UV behavior of the propagator of
the electric field must be $<E_i(x)E_i(y)>_{|x-y|<1/\mu}
\sim \frac{e^2}{|x-y|^2}$. Dimensional considerations, utilizing the
bosonized Hamiltonian \cite{pap} are consistent with this result.

We thus {\it assume} the following form of the UV asymptotics
of the correlators of the bosonic fields, which is the most general one
consistent with this UV scaling dimension, rotational symmetry,
and parity transformation properties,
\begin{eqnarray}
lim_{x\rightarrow y}<A_i(x)A_j(y)>&=&r_1\frac{\delta_{ij}}
{|x-y|}+2r_2\frac{(x-y)_i(x-y)_j}{|x-y|^3}; \\ \nonumber
lim_{x\rightarrow y}\frac{1}{e^2}<E_i(x)E_j(y)>&=&q_1
\frac{\delta_{ij}}
{|x-y|^2}+2q_2\frac{(x-y)_i(x-y)_j}{|x-y|^4}; \\ \nonumber
lim_{x\rightarrow y}<E_i(x)A_j(y)>&=&s_1\frac
{\delta_{ij}}{|x-y|^2}+2s_2\frac{(x-y)_i(x-y)_j}{|x-y|^4}\\ \nonumber
&+&mp_1\frac{\epsilon_{ij}}{|x-y|}+2mp_2\frac
{(\tilde x-\tilde y)_i(x-y)_j-
(\tilde x-\tilde y)_j(x-y)_i}{|x-y|^3}; \\ \nonumber
\label{correl}
\end{eqnarray}

The computational details are published elsewhere
\cite{pap}. Here, due to limitations of space, we only present the
results,
\begin{equation}
J_0\equiv\psi^\dagger\psi=\frac{1}{e}\partial_iE_i
\label{charge}
\end{equation}
\begin{equation}
J\equiv:\bar\psi\psi:=-2:\epsilon_{ij}A_iE_j:
\label{mass}
\end{equation}

The calculation of the spatial components of the current is more
involved. The result that
we get by expanding the defining relation eq.(\ref{bilinear}0
in powers of $1/\Lambda$ is,
\begin{equation}
J_i\equiv\bar\psi\gamma_i\psi=-e\kappa \Lambda A_i+
\frac{1}{e\Lambda}\left[
:\beta\tilde E_i(A\tilde E)+\gamma E^2A_i:\right]
\label{urrent}
\end{equation}

, where $E_i=\epsilon_{ij}E_j$.
The leading term is proportional to the UV cutoff. We
therefore have to keep also the next order term, which vanishes in
the naive continuum limit. (The cutoff-independent terms avearage to $0$).
The constants $\kappa$, $\beta$, and $\gamma$
depend on the regulator ratio $|\eta|/|\epsilon|$.
However, the final result should not depend on this ratio. This
means, that the formally small terms that we discarded do
not disappear completely, but renormalize the constants in
eq.(\ref{urrent})\footnote{One can show explicitly that the
higher order terms lead to a {\it finite} renormalization of the
coefficients \cite{pap}.}. The renormalized values of the
coefficients can be determined by requiring that the currents satisfy the
tree - level current algebra, possibly modified by Schwinger terms, and terms
that vanish in the continuum limit, and by the requirement of the restoration
of Lorentz invariance in the continuum limit. The situation
here is analogous to that in 1+1 dimensions
\cite{mandelstam}, where the use of Lorentz invariance is required
to fix the overall scale of the current.
The current is then determined as \cite{pap},
\begin{equation}
J_i\equiv\bar\psi\gamma_i\psi= - e\kappa\Lambda A_i+
\frac{1}{e\kappa\Lambda}\left[2\tilde E_i(A \tilde E)
- E^2A_i \right] + \frac{M}{e}\tilde E_i
\label{Current}
\end{equation}
The constants $\kappa$ and $M$ will be given below.
The fermionic bilinears are local
functions of $E_i$ and $A_i$. Their algebra can be calculated to give,
\begin{equation}
[J_i(x),J_j(y)]=2i\epsilon_{ij}J(x)\delta^2(x-y);\ \
[J_i(x),J(y)]=-2i\epsilon_{ij}J_j(x)\delta^2(x-y);
\label{algebra}
\end{equation}
$$[J_0(x),J_i(y)]=-i\left[\Lambda \kappa \delta_{ij}+
\frac{2}{e^2\Lambda\kappa}(E_iE_j-1/2E^2\delta_{ij})
\right]\partial^x_j\delta^2(x-y);$$
$$[J(x),J_0(y)]=\frac{2i}{e}E_i(x)\epsilon_{ij}\partial_j^x\delta^2(x-y)
$$
The first two commutators are the canonical tree - level ones.
The other two exhibit explicitly the Schwinger terms, which
appear also at the one loop level in perturbation theory.

We next calculate the bosonized form of the energy-momentum tensor of
the theory.  The gauge invariant, symmetric
energy-momentum tensor of QED$_3$ is given by,
\begin{equation}
T^{\mu\nu}=T^{\mu\nu}_B+T^{\mu\nu}_ F
\end{equation}
where the gauge and fermionic contributions are, respectively,
\begin{equation}
T^{\mu\nu}_B=
F^{\mu\lambda}F_{\lambda}^\nu+\frac{1}{4}g^{\mu\nu} F^2
\end{equation}
and,
\begin{equation}
T^{\mu\nu}_F=\frac{i}{4}(\bar\psi\gamma^\mu D^\nu\psi+\bar
\psi\gamma^\nu D^\mu\psi-D^\nu\bar\psi\gamma^\mu\psi
-D^\mu\bar\psi\gamma^\nu\psi)
\label{enmom}
\end{equation}
and where $D=\partial - ieA$ is the covariant derivative.

The general form of the bosonized Hamiltonian density following from
eq.(\ref{bilinear}) is,
\begin{equation}
T^{00}=\frac{1}{2}B^2+\frac{1}{2}E^2+
\frac{a}{e^2\Lambda}(\partial_iE_i)^2+b\frac{e^2}{4}\Lambda A^2+
\frac{1}{\Lambda}\left[c:(A\cdot\tilde E)^2:+
d:E^2A^2: \right]+fA\cdot\tilde E
\label{hamiltonian}
\end{equation}
If the theory is to be Lorentz invariant with standard transformation
properties for $E_i,B,J_0$, and $J_i$,
Maxwell's equations should be satisfied in the continuum limit. We therefore
require,
\begin{equation}
\dot B(x)=i[H, B(x)] =
-\epsilon_{ij}\partial_iE_j +O(1/\Lambda)
\end{equation}
\begin{equation}
\dot E_i=i[H, E_i]=-J_i+\epsilon_{ij}\partial_j B+O(1/\Lambda^2)
\end{equation}
Since $J_i$ itself contains terms of order $1/\Lambda$, we demand that the
second equation be satisfied to order $O(1/\Lambda^2)$.

It turns out  that this procedure still leaves the coefficients
$\kappa$ and $a$ undetermined. Our next step
will be therefore to impose the  complete Poincar\' e algebra.

A sufficient condition for Lorentz invariance of a theory invariant
under spatial rotations and translations
is the following commutation relation \cite{schwinger},
\begin{equation}
-i[T_{00}(x),T_{00}(y)]=-(T^0_i(x)+T^0_i(y))\partial^i\delta^2(x-y)
\label{schwinger}
\end{equation}
It can be easily verified (by multiplying eq.(\ref{schwinger})
by linear functions of $x$ and $y$ and integrating) that
eq.(\ref{schwinger})
implies the closure of the Poincar\' e algebra \cite{schwinger}, with
the following genertors:
the energy $\int d^2xT^{00}(x)$, the  momentum $P_i=\int d^2xT^0_i$,
and the angular momentum and boost generators, given, respectively by,
\begin{equation}
L_i=\int d^2x [x_0T^0_i-x_iT^{00}]; \ \ L=\int d^2x \epsilon_{ij}x_iT^0_j
\label{gen}
\end{equation}

We now require that the angular momentum generator has the standard
"orbital" and "spin" parts,
\begin{equation}
L\equiv\int d^2x
\epsilon_{ij}x_iT^0_j=\int d^2x[\epsilon_{ij}x_iE_k\partial_jA_k- A\cdot\tilde
E]
\end{equation}

Calculating the commutator in eq.(\ref{schwinger}), and performing the
operator product expansion using the forms of the propagators
in eq.(\ref{correl}, we find that Lorentz invariance is restored, in the
limit $\Lambda \rightarrow \infty$ for,

\begin{equation}
T^{00}=\frac{1}{2}B^2+\frac{1}{2}E^2+
\frac{1}{2e^2\kappa\Lambda}(\partial_iE_i)^2+\frac{e^2}{2}\kappa\Lambda A^2
\label{fin}
\end{equation}
$$
+\frac{1}{2\kappa\Lambda}\left[-2(A\cdot\tilde E)^2+
E^2A^2 \right]+M A\cdot\tilde E;
$$

\begin{equation}
T^{0i}= B\tilde E_i - (\partial_j E_j) A_i - \frac{2<B^2>}{\Lambda^3 \kappa}
\partial_i(AE) + \frac{M}{8} \partial_j(A_i \tilde A_j + A_j \tilde A_i);
\label{momentum}
\end{equation}

$$
J_i=-e\kappa\Lambda A_i+\frac{1}{e\kappa\Lambda}\left[2\tilde E_iA\cdot
\tilde E-A_iE^2\right]+\frac{M}{e} \tilde E_i
$$
with,
\begin{equation}
M=-\frac{2}{\kappa\Lambda}<A\cdot\tilde E>
\end{equation}
and $\kappa$ determined as,
\begin{equation}
\kappa=\frac{\Lambda}{e^2}\frac{<E^2>}{<B^2>}
\label{kappa}
\end{equation}

We have now determined all the coefficients in
the Hamiltonian density and the current
in terms of $\frac{<E^2>}{<B^2>}$ and $<A\cdot\tilde E>$.
We want to stress here that, although eq.(\ref{fin}) contains a parameter
$\kappa$, it does not define a one -  parameter set of theories.
The bosonized version of QED$_3$ corresponds to a unique choice
of $\kappa$, which satisfies eq.(\ref{kappa}). Unfortunately, since
eq.(\ref{fin}) defines a strongly interacting theory, we cannot determine
the numerical value of $\kappa$. This would involve the
solution of the model (at least in the UV region)
for arbitrary $\kappa$, calculating $<B^2>_\kappa$ and $<E^2>_\kappa$,
and solving the selfconsistency equation (\ref{kappa}).

We have therefore expressed completely the one flavor QED$_3$
in terms of the electric fields $E_i$, and their conjugate momenta
$A_i$. The resulting theory is local, and Lorentz invariant. The
fermionic bilinears satisfy a current algebra, which is modified
from its tree level form by Schwinger terms.

We now make several final comments.
First, the fermionic operators eq.(\ref{psi}) anticommute only
at distances larger than the ultraviolet cutoff. In fact, the
phase in the anticommutation relations contains a power tail
of the form $(1/(\Lambda |x-y|)^s$, where $s$ is a number of
order one. Therefore, also in the lattice version of this theory,
fermionic operators do not have canonical local anticommutation
relations.
This is the way in which the lattice theory avoids the standard
doubling problem in the continuum limit. This aspect is different
from the 1+1 dimensional case, where a theory of one staggered
lattice fermion leads in the continuum limit to a theory of one
Dirac fermion. Accordingly, the anticommutation relations in 1+1 dimensions
are canonical at all distances up to the cutoff.

Second, the fermionic operators we have constructed solve the
Gauss's constraint. If one relaxes this condition, there are additional
possibilities. One natural modification is to substitute for $E_i$
in eq.(\ref{vu}) the linear combination,
$\Pi_i=E_i+Ce^2 \epsilon_{ij}A_j$. This would give
$\psi^\dagger\psi=\frac{1}{e}\partial_iE_i+C e B$.
However, the coefficient $C$ in this case is not
arbitrary. It is crucial for the consistency of the derivation that
$\int d^2 x\partial_i\Pi_i$ has quantized eigenvalues
\cite{pap}. Since the magnetic flux in $QED_3$ is quantized
in units of $\frac{2\pi}{e}$, the coefficient $\kappa$ must
be given by $\frac{n}{2\pi}$ with integer $n$. This
possible modification is a reflection of  the well known
regularization ambiguity in fermionic QED$_3$ \cite{coste},
which leads to the appearance of an induced Chern - Simons term
in the action.

Third, although the theory we have derived is Lorentz invariant, it
is not written in terms of Lorentz covariant fields.
The appearance of the UV cutoff in the expressions for
the current and the Hamiltonian also point to the fact that the
variables $E_i$ and $A_i$ are not the most convenient ones. It
would be desirable to make one more step, and find a formulation
in terms of a different, covariant bosonic field. We believe, that
the most natural candidate for this field is the magnetic vortex field.
In 2+1 dimensions it is a
scalar \cite{npb}. When written in terms of this field, the electric
current is  trivially conserved, and
the electric charge has the interpretation of a topological winding
number (which is quantized classically) \cite{review}.

Finally, we note that the fermionic operators
which we have constructed involve a {\it product}
of vortex and antivortex operators, carrying opposite magnetic fluxes.
Therefore the operators do not carry net magnetic flux, and thus are
{\it not} analogous to $2+1$ - dimensional dyons. The mechanism of their
anticommutativity is different from that discussed in \cite{wilczek}, and
subsequently extensively exploited in the analysis of Chern - Simons theories.
In fact, our construction has a natural generalization to $3+1$ dimensions,
where the role of the vortex - antivortex pair is taken over by
an infinitesimal closed t'Hooft loop of half - integer
strength. Work along these lines is currently in progress.

{\bf Acknowledgements.} We thank T. Jaroszewicz for discussions.
We are indebted to Y. Kluger and especially to T. Bhattacharya
for numerous discussions and many helpful suggestions.

\end{document}